\documentclass[aps,twocolumn,showpacs,floats,pre]{revtex4}

\usepackage{amsfonts}
\usepackage{amsmath}
\usepackage{amssymb}
\usepackage{amstext}
\usepackage{bm}

\usepackage{algorithm,algorithmic}

\usepackage{epsf,graphicx}
\graphicspath{{./Figures/}}

\usepackage{url}          

\begin{document}


\title{Mean field initialization of the Annealed Importance Sampling algorithm for
an efficient evaluation of the Partition Function of Restricted Boltzmann Machines}




\author{A. Prat Pou$^{\rm a}$, E. Romero$^{\rm b}$ J. Mart\'i$^{\rm a}$, F. Mazzanti$^{\rm a}$}

\affiliation{
 $^{\rm a}$ Departament de F\'{i}sica,
  Universitat Polit\`{e}cnica de Catalunya, Barcelona Tech, 
  Campus Nord B4-B5, E-08034, Barcelona, Spain \\
 $^{\rm b}$ Departament de Ci\`encies de la Computaci\'o, 
  Universitat Polit\`{e}cnica de Catalunya, Barcelona Tech, 
  Campus Nord B4-B5, E-08034, Barcelona, Spain
}


\begin{abstract}
  
  Probabilistic models in physics often require from the evaluation of
  normalized Boltzmann factors, which in turn implies the computation of the
  partition function $Z$. Getting the exact value of $Z$, though, becomes a
  forbiddingly expensive task as the system size increases. This problem is
  also present in probabilistic learning models such as the Restricted
  Boltzmann Machine~(RBM), where the situation is even worse as the exact
  learning rules implies the computation of $Z$ at each iteration. A possible
  way to tackle this problem is to use the Annealed Importance Sampling~
  (AIS) algorithm, which provides a tool to stochastically estimate the
  partition function of the system. So far, the standard application of the
  AIS algorithm starts from the uniform probability distribution and uses a
  large number of Monte Carlo steps to obtain reliable estimations of $Z$
  following an annealing process. In this work we show that both the quality
  of the estimation and the cost of the computation can be significantly
  improved by using a properly selected mean-field starting probability
  distribution. We perform a systematic analysis of AIS in both small- and
  large-sized problems, and compare the results to exact values  in problems
  where these are known. As a result of our systematic analysis, we propose
  two successful strategies that work well in all the problems analyzed. We
  conclude that these are good starting points to estimate the partition
  function with AIS with a relatively low computational cost.
  
\end{abstract}


\maketitle

\section{Introduction}

The evaluation of thermodynamic potentials such as the entropy or free energy
is key to understand the equilibrium properties of physical 
systems~\cite{Callen_1985}. In real-sized
classical problems, computer simulations based on Molecular Dynamics or Monte
Carlo methods can not generically access them mainly because of the size of
the spaces of states to sample, which grows exponentially with the number of
particles. This effect is particularly easy to quantify in magnetic models of
classical two-state spin systems, where the volume of the phase space grows
as $2^N$ with $N$ the total number of spins. Quantities such as the
Helmholtz free energy $F$ in the canonical ensemble, proportional to the
logarithm of the partition function~\cite{Huang_1987, Pathria_1998}
\begin{equation}
  Z = \sum_{{\bf x}} e^{-E({\bf x})/k_B T} \ ,
  \label{Z}
\end{equation}
are out of reach as the sum extends over all possible states ${\bf x}$, with   
$E({\bf x})$ the corresponding energy, $k_B$ the Boltzmann's constant, and  
$T$ the temperature. Actually, finding the value of $Z$ is known to be an
NP-hard problem~\cite{Goldberg_15} that therefore prevents an exact estimation
unless the system is small. This same quantity, with the same problems, appears
also in other areas of science, as for instance in machine learning models
based on Restricted Boltzmann Machines{\cite{Smolensky_1986, Bengio_09}. In
that case the situation is even worse because $Z$, which is the normalization
factor of a Gibbs probability distribution, has to be evaluated at each
iteration in a learning scheme if the gradients of the cost function are to be
estimated exactly.

The relevance but unfortunate computational complexity implied in the
determination of $Z$ has raised the urge to devise methods to approximate it
in a tractable way. One remarkable technique designed to tackle this problem
was developed by Bennett~\cite{Bennet_1976}, where the free energy difference
between two overlapping canonical ensemble is estimated directly in a Monte
Carlo simulation. In case one of the two values of $F$ is known, the method
allows obtain the value of the other, thus gaining access to
$F= -k_B T \log(Z)$. Another interesting approach towards the evaluation of the
partition function is derived from the Wang-Landau 
algorithm~\cite{Landau_2004, Zhou_2006, Vogel_2013}, 
where a stochastic exploration of the phase space is used to recover the
density of energy states $\rho(E)$ corresponding to the Hamiltonian of the
system under study. In this framework, the partition function is recovered as
the integral of $\rho(E) e^{-E/k_B T}$ over the energy range spanned by the
system configurations. This method has proved to reliably reproduce the
physics of different systems as the 2D-Ising model, although it can be
difficult to apply to more complex situations involving an intricated 
$\rho(E)$.

An alternative approach to the problem was devised in 2001 by 
R.~M.~Neal~\cite{Neal_1998, Neal_2001}, the Annealed Importance Sampling algorithm, 
where an annealing procedure is implemented to obtain reliable samples from an
otherwise intractable probability distribution starting from samples of a
simpler and tractable one. In this method, the partition function is one of
the simplest quantities to evaluate, although as in most sampling schemes,
convergence towards the exact value of $Z$ is only guaranteed in the infinite
limit, both in number of samples and intermediate annealing steps. In
practical terms, when a finite number of samples and intermediate annealing
chains is employed, the predicted value of $Z$ depends on the different
simulation inputs, particularly on the initial probability distribution.

Surprisingly, and despite its broad formulation in terms of an initial and a
final probability distributions, little use has been seen of the AIS
algorithm in the numerical simulation of physical systems to the best of our
knowledge. More application has emerged in the world of neural networks,
particularly in the field of RBMs~\cite{Smolensky_1986}, where the evaluation
of $Z$ is key to a precise optimization of the system parameters along
learning. In this context, the AIS algorithm turns out to be most efficient
since the random walk exploration can be performed by means of Gibbs
sampling, which is fully parallelizable~\cite{Salak_08}. Finally, the AIS
algorithm is particularly suited to address binary state unit problems like
spin systems or RBMs where the different probability distributions involved
along the annealing chains are cost-effective and simple to evaluate.

In this work we study how AIS can be used to produce reliable estimates of 
$\log(Z)$ in RBMs and systems that can be mapped to them. Our goal is to 
achieve that with a small computational cost, even in realistically large
problems, when a suitable starting probability distribution is employed. We
discuss how to obtain the optimal mean field probability distribution 
$p_0^*({\bf x})$ that is closest to the Boltzmann distribution of the real
model under study. After a brief derivation of how to get $p_0^*({\bf x})$
from average system properties, we propose two strategies to find
approximations to it in both artificial sets and magnetic spin systems where
the exact value of the partition function can be determined. Finally, we
compare the results obtained with the standard procedure, where the uniform
probability distribution is employed as the starting point of the AIS
algorithm.

\section{Annealed Importance Sampling}
\label{section-AIS}


The AIS algorithm, developed by R.~Neal in the late
90's~\cite{Neal_1998, Neal_2001}
allows sampling from a probability distribution that would otherwise be
intractable. It can be used to estimate $Z$, but it is more general and
allows finding approximate values of any observable quantity
$\alpha({\bf x})$ over a probability distribution $p({\bf x})$. In a general
sense, this computation can be very inefficient due to two main reasons. On
one hand, the probability distribution $p({\bf x})$ can be impossible to
sample because the exact form of $p({\bf x})$ is not known, as it happens
in many quantum physics problems
\cite{Ceperley_1996, Kosztin_1996, Sarsa_2000}
On the other hand, the number of samples required to obtain an accurate
estimate of the average value of $\alpha({\bf x})$ may be unreasonably large.
In order to deal with these problems, one usually resorts to some form of
Importance Sampling, where the exploration of the space is guided by a known
and suitable probability distribution $q({\bf x})$\cite{Srinivasan_2002}.
In this way one typically evaluates
\begin{equation}
  \langle \alpha \rangle = \int d{\bf x}\,q({\bf x})
  \left(
  { p({\bf x}) \alpha({\bf x}) \over q({\bf x}) } 
  \right) \ .
  \label{xi-is}
\end{equation}
using stochastic techniques, where samples are drawn from $q({\bf x})$.
Importance Sampling is employed to reduce the variance of the estimator, or to
reduce the number of samplings needed to achieve the same statistical
accuracy.
In any case, Importance Sampling can only be performed when a suitable
$q({\bf x})$ is at hand, but that may not always be the case. The AIS
method allows building
$q({\bf x})$ starting from a trivial probability distribution, and performing
an annealing process through a set of intermediate distribution corresponding
to decreasing temperatures.


As explained in~\cite{Neal_1998, Neal_2001},
in order to estimate $\langle\alpha\rangle$ 
starting from a trivial $p_0({\bf x})$, 
one builds a chain of intermediate distributions $p_i({\bf x})$ that
interpolate between $p_0({\bf x})$ and 
$p_n({\bf  x})=p({\bf x})$.
Denoting by $\tilde p_k({\bf x})=Z_k p_k({\bf x})$ the corresponding
unnormalized probability distributions, a
common scheme is to define
\begin{equation}
  \tilde p_k({\bf x}) = \tilde p_0({\bf x})^{1-\beta_k}
  \tilde p_n({\bf x})^{\beta_k}
  \ ,
\label{p_AIS}
\end{equation}
with $0 = \beta_0 < \beta_1 < \cdots < \beta_n = 1$
and $N_\beta=n+1$. 
The approach used in AIS is to turn the estimation of $\langle \alpha\rangle$ into a
multidimensional integration of the form


\begin{equation}
  \langle\alpha\rangle = \int d{\bf x}_1 \cdots d{\bf x}_n \,
  g({\bf x}_1, \cdots, {\bf x}_n)
                    { f({\bf x}_1, \cdots, {\bf x}_n) \over
                      g({\bf x}_1, \cdots, {\bf x}_n) }
                    \alpha({\bf x}_n)
                    \ ,
  \label{xi_AIS}
\end{equation}
where
\begin{eqnarray}
  f({\bf x}_1, \cdots, {\bf x}_n) \!\!\! & = & \!\!\! p_n({\bf x}_n)
  \prod_{j=1}^{n-1}
  \hat T_j({\bf x}_{j+1}, {\bf x}_j)
  \label{fn} \\
  g({\bf x}_1, \cdots, {\bf x}_n) \!\!\! & = & \!\!\! p_0({\bf x}_1)
  \prod_{j=1}^{n-1}
  T_j({\bf x}_j, {\bf x}_{j+1})  
\end{eqnarray}
are normalized joint probability distributions for the set of
variables $\{ {\bf x}_1, \ldots, {\bf x}_n\}$.
In these expressions $T_k({\bf x},{\bf y})$ represents a transition
probability of moving from state ${\bf x}$ to state ${\bf y}$, which
asymptotically leads to the equilibrium probability $p_k({\bf x})$.
In the same way, $\hat T_k({\bf y},{\bf x})$ represents the reversal
of $T_k({\bf x},{\bf y})$.  The detailed balance condition implies
that the transition probabilities fulfill the relation
\begin{equation}
  \hat T_j({\bf y}, {\bf x}) = T_j({\bf x}, {\bf y})
         {p_j({\bf x}) \over p_j({\bf y})} 
  \label{det_balance}
\end{equation}  
in order to be able to sample the space ergodically~\cite{Amit_89}.
Therefore, $\langle\alpha\rangle$ can be estimated from Eq.~(\ref{xi_AIS})
with 
\begin{equation}
  { f({\bf x}_1, \ldots, {\bf x}_n)
    \over
    g({\bf x}_1, \ldots, {\bf x}_n)
  } =
  \prod_{k=1}^n { p_k({\bf x}_k) \over p_{k-1}({\bf x}_k) } \ ,
  \label{fg_1}
\end{equation}
as $g({\bf x}_1, \ldots, {\bf x}_n)$ is easily sampled from the trivial $p_0({\bf x})$.

In practice, one uses $g({\bf x}_1, \ldots, {\bf x}_n)$ to generate
$N_s$ samples of all the intermediate distributions, such that for
every set of values $\{ {\bf x}_1^i, {\bf x}_2^i, \ldots, {\bf x}_n^i
\}$,
with $i=1,2, \ldots, N_s$,
one gets a set of weights $\{\omega_i\}$ upon substitution in
Eq.~(\ref{fg_1}). In this way, $\langle \alpha\rangle$ is estimated
according to
\begin{equation}
  \langle \alpha\rangle \approx
          { \sum_{i=1}^{N_s} \omega_i \alpha({\bf x}_n^i) \over
            \sum_{i=1}^{N_s} \omega_i } \ ,
  \label{xi_AIS_b}
\end{equation}
with
\begin{equation}
\omega_i = \prod_{k=1}^n { p_k({\bf x}^i_k) \over p_{k-1}({\bf x}^i_k)} =
        {Z_0 \over Z_n}
    \prod_{k=1}^n { \tilde p_k({\bf x}^i_k) \over \tilde p_{k-1}({\bf
        x}^i_k) } = {Z_0 \over Z_n} \tilde\omega_i \ ,
\label{omegai_k}
\end{equation}
which defines the set of importance weights $\{\tilde\omega_i\}$
obtained from the product of the ratios of the unnormalized
probabilities. Notice that $\tilde \omega_i$ is an accessible quantity,
while $\omega_i$ is not, just because one does not have access to $Z_n$.
One important consequence of this formalism is that a simple estimator
of the partition function $Z_n$ associated to the distribution
$p_n({\bf x})=p({\bf x})$ is directly given by the average value
\begin{equation}
  {Z_n \over Z_0} \approx {1\over N_s} \sum_i \tilde\omega_i \ .
  \label{Z_AIS_wi}
\end{equation}

Since the values of $\tilde\omega_i$ are usually large,  one
typically draws samples of $\log(\tilde\omega_i)$.
%
In this way, one uses a set of
$Z_0$-normalized AIS samples $s_i = \log(\tilde\omega_i)+\log(Z_0)$, such that
\begin{equation}
\log(Z_n) \approx
\log\left[{1\over N_s} \sum_i e^{s_i}\right] = 
\log(Z_{\rm AIS}) \ ,
\label{log_mean_exp}
\end{equation}
and defines $Z_{\rm AIS}$ as an approximation to $Z_n$. Notice that
this value is different from the mean of the samples $s_i$, although
these two quantities do not differ much when the variance of the samples
is small.
In fact, these two quantities tend to be the
same when the variance of the set of samples is small compared to the
mean value. In other situations, the nonlinear character of the
operation in Eq.~(\ref{log_mean_exp}) makes the result be dominated by
the largest samples, to the point that, in the extreme case, the
largest sample exhausts the total sum.

\section{The Restricted Boltzmann Machine}
\label{sec_RBM}

A RBM with binary units is a spin model describing a mixture of two different
species, where intra-species interactions are forbidden, and units play the
role of the spins. In general, though, RBM units take $[0,1]$ values rather
than $[-1,1]$. Furthermore, only one component of this mixture is assumed to
be accessible to the external observer, usually called the {\it visible 
layer}. The other species, usually called {\it hidden layer}, is assumed to
have no contact with the outside world, and is present to build up
correlations in the model. As a consequence, one is only interested in the
marginal probability distribution associated to the visible units.

The energy function of a binary RBM with $N_v$ visible units ${\bf x}$ 
and $N_h$ hidden units ${\bf h}$, is defined as~\cite{Hinton_2006, Salak_2007}:
\begin{equation}
 E({\bf x},{\bf h}) = -{\bf x}^{\rm T}{\bf b} - {\bf c}^{\rm T}{\bf h} -
 {\bf x}^{\rm T}{\bf W}{\bf h} \ ,
\label{energy-RBM}
\end{equation}
where ${\bf W}$
is the two-body weights matrix setting the 
coupling strength between the two species, while $\bf b$ and $\bf c$
represent the external fields acting on each layer and are generically
denoted as {\it bias}. In this expression, ${\bf x}^{\rm T}$ stands for the 
transpose of vector ${\bf x}$.

The energy in Eq.~(\ref{energy-RBM}) 
can be cast as a quadratic form, where visible and
hidden units are organized as row and column vectors preceded by a
constant value of $1$ to account for the bias terms
\begin{equation}
  \tilde {\bf x}^{\rm T} = 
  (1 \,x_1\, x_2 \cdots x_{N_v})
  \,\,\,\,\, ,  \,\,\,\,\,
  \tilde {\bf h}^{\rm T} =
    (1 \,h_1\, h_2 \cdots h_{N_h}) \ ,
\end{equation}
leading to
\begin{equation}
  E(\tilde{\bf x}, \tilde{\bf h}) = -\tilde{\bf x}^{\rm T}
  \left(
  \begin{array}{cc}
    0       & {\bf c}^{\rm T} \\
    {\bf b} & {\bf W} 
  \end{array}
  \right) \tilde{\bf h} 
  \equiv
 -\tilde{\bf x}^{\rm T} \tilde{\bf W} \tilde{\bf h}\ ,
\label{energy-RBM-mat}
\end{equation}
where $\tilde {\bf W}$ is the {\em extended} weights matrix, which 
includes the bias terms.

As usual in energy-based models, the probability of a state $({\bf x},{\bf h})$ 
follows a Boltzmann distribution
\begin{equation}
\label{probability-RBM-xh}
 p({\bf x},{\bf h}) = \frac{e^{-E({\bf x},{\bf h})/T}}{Z} \ ,
\end{equation}
with 
\begin{equation}
\label{partfun-RBM-1}
  Z = \sum_{{\bf x},{\bf h}} {e^{-E({\bf x},{\bf h})/T}}
\end{equation}
and $k_B$ set to 1.
The particular form of the energy function (\ref{energy-RBM}) makes
both $P({\bf h}|{\bf x})$ and $P({\bf x}|{\bf h})$ to factorize
as a product of probabilities corresponding to independent random
variables. As a consequence, Gibbs sampling can be efficiently used to compute
them~\cite{Geman_1984}. In addition,
it is also possible to evaluate one of the two sums involved in
the partition function
In this way, for $[0,1]$ units, one has
\begin{equation}
\label{partfun-RBM-2}
Z = \sum_{\bf x} e^{{\bf x}^{\rm T}{\bf b}/T} 
    \prod_j \left(1 + e^{({\bf c}_j + {\bf x}^{\rm T}{\bf W}_{\!j})/T}\right),
\end{equation}
where index $j$ runs over the whole set of hidden units, and 
${\bf W}_{\!j}$ stands for the $j$th column of ${\bf W}$. 
However, the evaluation of
$Z$ is still prohibitive when the number of input and
hidden variables is large, since it involves an exponentially large
number of terms. 
For that reason, RBMs are computationally hard to evaluate or 
simulate accurately~\cite{Long_2010}.

\section{Datasets}
\label{sec_Datasets}

In this work we explore different problems where 
$\log(Z)$ can be exactly computed, which will be then used to 
benchmark the approximations described afterwards.
At the end, these are employed to predict 
the value of $\log(Z)$ on a large, realistic system where 
an exact evaluation is prohibitive. 
The set of models where the exact $\log(Z)$ is accessible include
artificially generated weights, RBM learning weights, 
and magnetic spin systems that can be directly mapped into an RBM.
These include:



\begin{itemize}

\item[1)] Gaussian Weights with Gaussian Moments (GWGM), 
  characterized by an extended matrix of weights $\tilde {\bf W}$ of Gaussian random
  numbers with $N_v=20$ and $N_h=180$. The mean value and the standard
  deviation of each set of weights are 
  also sampled from a Gaussian distribution
  with $\mu=-10, \sigma=10$, and $\mu=20, \sigma=10$, respectively. The 
  temperature for all these models has been set to 1.
  Due to the reduced value of $N_v$, the exact calculation of $Z$ can be
  performed by brute force, and we have generated a total of 100 models.

\item[2)] A set of weights obtained after training a RBM with the
  MNIST dataset~\cite{LeCun-http}, with 
  $N_h=20$ hidden units (MNIST-20h), similar to the simple case
  studied in Ref.~\cite{Salak_08} 
  We monitor and store the weights
  along the learning process
  with the aim of having a complete picture of their evolution.
%
%
  In this way, we have snapshots 
  taken at the beginning of the learning, where the training set
  typically does not correspond to the highest probability states, and
  at the end, where they are supposed to carry most of the probability
  mass.
  As in the GWGM dataset, $T$ is set to 1.

\end{itemize}
These two datasets use $[0,1]$ binary visible and hidden variables.


\begin{itemize}
  \item[3)] Classical Ising and Spin Glass models in one and two dimensions.
    A one-dimensional Ising model with periodic boundary conditions containing an 
    even number of spins $\{s_1,s_2,\dots,s_{2N}\}$ can be represented by a RBM with 
    the same number of units in each layer. Identifying even and odd spins with 
    hidden and visible units, respectively, one has
    \begin{eqnarray*}
     {\bf b}^{\rm T} \!\!\! & = & \!\!\! (B_1, B_3, \cdots, B_{2N-1}) \\
     {\bf c}^{\rm T} \!\!\! & = & \!\!\! (B_2, B_4, \cdots, B_{2N})
    \end{eqnarray*}
    and 
    \begin{eqnarray*}
    {\bf W} \!\!\! & = & \!\!\!
    \left(
    \begin{array}{lllll}
      J_{1,2}  & 0       & 0      & \cdots & J_{N,1} \\
      J_{2,3}  & J_{3,4} & 0      & \cdots & 0      \\
      0       & J_{4,5}  & J_{5,6} & \cdots & 0      \\
      \vdots  & \vdots  & \vdots & \ddots & \vdots \\
      0       & 0       & 0      & \cdots & J_{N-1,N}
    \end{array}
    \!\!\right),
    \end{eqnarray*}
    where $J_{i,i+1}$ is the interaction between spins $s_i$ and
    $s_{i+1}$. Only two entries per row/column can be non-zero in this arrangement.
    In the Ising model (1DIsing), $J_{i,i+1} = J$ and $B_i = B$ for all spins, while 
    they can take different values in what we denote as a Spin Glass model (1DSG).
    The partition function of 1DIsing and 1DSpinGlass can
    be easily computed using the Transfer Matrix formalism~\cite{Ising_1925, Binney_92}.
    We have generated 3 sets of 100 1DIsing models, as well as 3 sets of 100 1DSG models, all
    of them containing $N_s=200$ spins. In both cases (Ising and Spin Glass) the 
    $J_{i,j}$ and $B_i$ Hamiltonian parameters have been drawn at random from 
    a Normal distribution 
    with $\mu=-100$ and $\sigma=200$, for three different temperatures
    $T=10$ (1DIsing1 and 1DSG1), $T=1$ (1DIsing2 and 1DSG2), and 
    $T=0.1$ (1DIsing3 and 1DSG3).

    The two-dimensional square-lattice Ising model is much harder to solve and
    its analytic solution was given by Onsager in~\cite{Onsager_1944} in the 
    absence of an external field. 
    Similar to the 1D models, it can be represented by an RBM, 
    where visible and hidden units are arranged in a checkerboard configuration.
    In this case, four weights can be non-zero in each row and
    column of $\tilde {\bf W}$ since there are no bias terms. 
    Three sets of 100 2DIsing models (2DIsing1, 2DIsing2 and 2DIsing3)
    corresponding to $N_s=256$ spins have been generated, 
    with parameters drawn from the same normal distributions used for the
    previous 1D cases, and the same temperatures.

    Furthermore, we have extend that to what we name a 2D Spin Glass (2DSG),
    where all two-body $J_{i,j}$ correlations are different, while keeping the 
    connectivity restricted to nearest neighbors. 
    In this case the partition function is computed by {\em brute force},
    which limits the size of the square lattices to be less than or equal to $6\times 6$,
    as an even number of spins per dimension is required in order to properly satisfy
    the periodic boundary conditions. Two different sets of 50 models (2DSG1 and 2DSG2) 
    have been used, drawn from a normal distribution with $\mu=10^4$ and $\sigma=10^4$ 
    and corresponding to $T=10$ and $T=0.1$, respectively.


\end{itemize}
All these models use $[-1,1]$ spin variables as standard.


Finally, we also analyze the MNIST dataset with a RBM containing $N_h=500$ hidden
units (MNIST-500h), where no exact value of $\log(Z)$ is known. As in the 
MNIST-20h, they are the result of a learning process with $T=1$ and $[0,1]$ units.



\section{The optimal mean field approximation}
\label{sec_Boptimal}

The equilibrium Boltzmann distribution associated to any physical
system is given by
\begin{equation}
  p({\bf x}) =  
  {e^{-E({\bf x})/T} \over Z} \ ,
  \label{Boltz_RBM}
\end{equation}
where $E({\bf x})$ is the system's energy corresponding to state ${\bf x}$.
In the spirit of AIS, the partition function associated to $p_n({\bf x})$ 
can be obtained from a chain of
intermediate probability distributions that start from another, much simpler 
and easy to sample 
$p_0({\bf x})$, as shown in Sec.~\ref{section-AIS}. Getting a good $p_0({\bf x})$ 
can easen the job to AIS, and becomes therefore a key ingredient to
get an accurate estimation of $\log(Z)$ with a reasonable number of
intermediate chains and samples. 
A very simple  probability distribution $p_0({\bf x})$ can be obtained from a
mean-field model containing only external fields ${\bf B}$.
%
%
%
In this scheme, and for an RBM, $E_0({\bf x})=-{\bf x}^{\rm T} {\bf B}$ defines the
starting mean field energy, which makes 
\begin{equation}
p_0({\bf x}) = {2^{N_h}e^{{\bf x}^{\rm T}\cdot{\bf B}/T} \over Z_0}
\label{p0_meanfield}
\end{equation}
be the product of independent distributions for each unit,
thus allowing for a very simple and efficient sampling scheme in parallel.
Furthermore, for $[0,1]$ binary units, the corresponding partition function reads
\begin{equation}
Z_0 = 2^{N_h} \prod_{j=1}^{N_v} (1 + e^{B_j/T})
\label{logZ0_01}
\end{equation}
while for $[-1,1]$ units one has
\begin{equation}
Z_0 = 2^{N_v+N_h} \prod_{j=1}^{N_v} \cosh(B_j/T) \ .
\label{logZ0_pm1}
\end{equation}

Despite dealing with a mean field, getting the most suitable ${\bf B}$ may not
be a trivial task.
In most practical
applications, and with lack of a better model, the simplest choice 
${\bf B}=0$ is adopted, thus turning $p_0({\bf x})$ into the uniform
probability distribution. In the spirit of the AIS algorithm, and according
to the theoretical development~\cite{Neal_1998,Neal_2001}, one then expects
that increasing the number of intermediate distributions should lead to the
exact result, no matter what the starting $p_0({\bf x})$ is. Whilst this
should be the case, it is not clear how the dynamics of this process is, and
whether the desired limit is attained with a large but manageable number of
intermediate distributions. In other words, one has no clue as to what the
convergence properties of the algorithm are, other than knowing that it
provides the right result in the infinite limit. In order to test that in
practice, we have conducted different experiments with the GWGM and MNIST-20h
datasets of Sec.~\ref{sec_Datasets}. Figure~\ref{fig_conv_B0_MNIST_Lokos} 
shows the evolution of the prediction of $\log(Z)$ with $N_\beta$ for the 
MNIST-20h (left panel) and 10 randomly selected
GWGM weights (right panel). In all these calculations, a total of $N_s=1024$ 
AIS samples have been employed to build $\log(Z_{AIS})$ according 
to Eq.~(\ref{log_mean_exp}). In the MNIST-20h, both the exact and the
predicted values are displayed, while in the GWGM case the ratio of the AIS
$\log(Z)$ to the exact $\log(Z)$ is displayed for the sake of clarity. The
errorbars are obtained after averaging 100 repetitions of the same
experiments.

\begin{center}
\begin{figure}[!t]
\includegraphics[width=\linewidth]{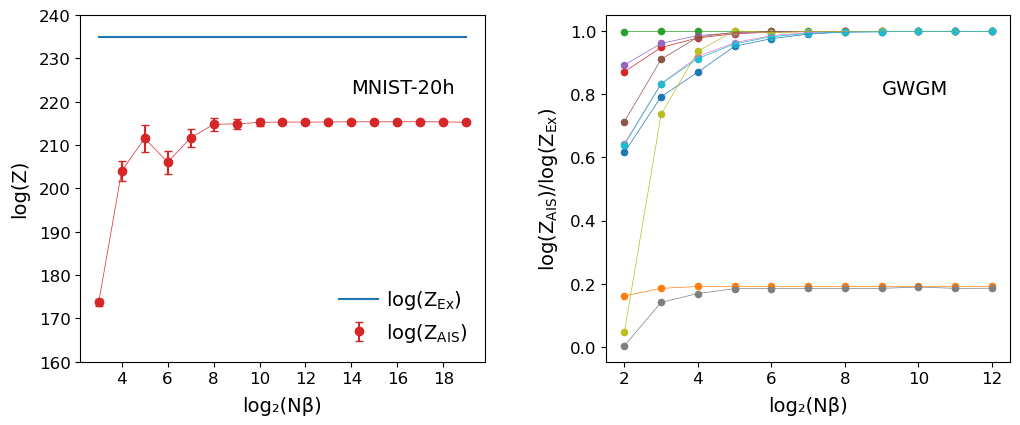}
\caption{AIS estimation of $\log(Z)$ starting from ${\bf B}=0$ for the
 MNIST-20h (left) and ten different GWGM datasets (right) as a function of
 the number $N_\beta$ of intermediate distributions. The left panel shows
 both the exact value(in blue) and the AIS estimations, while on the right
 the ratio of these two quantities is plotted.}
\label{fig_conv_B0_MNIST_Lokos}
\end{figure}
\end{center}

Two immediate conclusions can be drawn from Fig.~\ref{fig_conv_B0_MNIST_Lokos}. 
On one hand, it is clear that in both cases a stable prediction has been
achieved already at $N_\beta=N_s=1024$. This 
fact has also been observed 
with the other datasets tested. Starting from there, we have set
$N_\beta=4096$ and $N_s=1024$ in all the following AIS runs all over this
work, which seems to be large enough to get stable results while still
allowing for a fast evaluation of $\log(Z)$ with a standard computer. On the
other hand, one readily notices that, despite providing an apparently
converged result, the AIS prediction starting from ${\bf B}=0$ may differ
substantially from the exact result, even in cases where one of the
dimensions of the problem ($N_v$ or $N_h$) is small. The situation is even
worse as the errorbars diminish with increasing $N_\beta$, leading to the
false impression that a reliable prediction has been achieved. The results in
the left panel show that this picture remains unaltered even with $N_\beta =
2^{20}$, thus indicating that probably a completely unpractical amount of
intermediate distributions is needed to produce the required changes to bring
the AIS prediction close to the exact result, something that is guaranteed in
the asymptotic limit~\cite{Neal_1998,Neal_2001}.

Still, the plots in Fig.~\ref{fig_conv_B0_MNIST_Lokos} yield a discouraging
picture about the possibility of achieving good results starting from ${\bf B}=0$, 
an image that should be properly put into perspective. In order
to get a more complete view we have conducted AIS experiments starting from ${\bf B}=0$ 
on all the datasets of Section~\ref{sec_Datasets}. We have
computed 100 independent repetitions in all cases, consisting each in
$N_s=1024$ AIS samples with $N_\beta=4096$.
For every model, an estimation of $\log(Z)$ has been obtained from the 1024 samples
using Eq.(\ref{log_mean_exp}), and the relative error
\begin{equation}
  \epsilon_r =
  \left|
  {\log(Z_{\rm Ex})-\log(Z_{\rm AIS}) \over \log(Z_{\rm Ex}) }
  \right| \ ,
\label{relative_error}
\end{equation}
has been subsequently computed and averaged over all models belonging to the
same dataset. The result is shown in Fig.~\ref{fig_Percentage_Beq0}, and
displays the percentage of samples fulfilling the condition
$\epsilon_r\leq0.05$. As it can be seen, the choice ${\bf B}=0$ works
in many cases, but not in all of them.

\begin{center}
\begin{figure}[!t]
\includegraphics[width=0.9\linewidth]{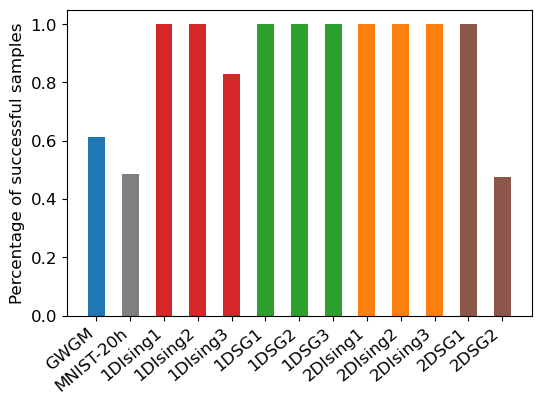}
\caption{Percentage of AIS samples producing an estimation of $\log(Z)$ with a relative error 
of less that $5\%$ with respect to the exact result, obtained starting from ${\bf B}=0$.
The results have been averaged over all models of each tested dataset.}
\label{fig_Percentage_Beq0}
\end{figure}
\end{center}

In any case, and despite the fact that the uniform probability distribution
corresponding to ${\bf B}=0$ provides a trivial starting point, it is not the
only possible simple choice. In fact, any distribution of the mean field form
given in Eq.~(\ref{p0_meanfield}) is suitable to start AIS from, as with
that
all components of ${\bf x}$ become independent random variables that can be
sampled in parallel. Among all the possible choices of $
{\bf B}$, therefore, one can look for the optimal one that produces the best
possible results with little computational cost. In this context, being optimal
means to produce a mean field probability distribution that is closest to the
actual $p_n({\bf x})$ one seeks to sample, according to some metric.

In particular, the optimal values ${\bf B}^*$ of ${\bf B}$ can be obtained minimizing the
Kullback-Leibler (KL) divergence between $p_0({\bf x})$ and the full RBM
probability distribution $p_n({\bf x})$, so we impose the condition
\[
\left.
\nabla_{\bf B} \sum_{\bf x} p_n({\bf x}) \log\left( p_n({\bf x}) \over p_0({\bf x}) 
\right) 
\right|_{{\bf B}={\bf B}^*}
= 0 \ ,
\]
where the sum over ${\bf x}$ extends to all the $2^{Nv}$ states as hidden states 
have already been marginalized in both $p_0({\bf x})$ and $p_n({\bf x})$.
One thus have, for $x_i\in[0,1]$
\begin{eqnarray}
0 & = & 
\left.
-\sum_{\bf x} p_n({\bf x}) \nabla_{\bf B} \log p_0({\bf x})
\right|_{{\bf B}={\bf B}^*}
\nonumber \\
& = & -{1\over T}\sum_{\bf x} p_n({\bf x}) {\bf x} + 
\left. 
\sum_{\bf x} p_n({\bf x}) \nabla_{\bf B} \log Z_0 \right|_{{\bf B}={\bf B}^*}
\nonumber \\
& = & -\langle {\bf x} \rangle_n +
{1 \over 1 + e^{-{\bf B}^*/T}} \ ,
\end{eqnarray}
where the subscript $n$ indicates that the average values are taken
over the $p_n({\bf x})$ probability distribution corresponding to the
target RBM. In this way one gets, for $x_i\in[0,1]$
\begin{equation}
{\bf B}^* = -T\log\left( {1\over \langle {\bf x} \rangle_n} - 1 \right) 
\label{Bi_xin}
\end{equation}

for each visible unit $i\in \{1,2,\ldots, N_v\}$. For $x_i\in[-1,1]$, 
a similar procedure leads to
\begin{equation}
{\bf B}^* = -T\tanh^{-1}(\langle {\bf x} \rangle_n) \ .
\label{Bi_xin_pm1}
\end{equation}
These expressions imply that the problem of finding 
${\bf B}^*$ is 
equivalent to obtaining the exact average values of the visible units, which
may not be a trivial task depending on the problem at hand. 

\begin{center}
\begin{figure}[!t]
\includegraphics[width=\linewidth]{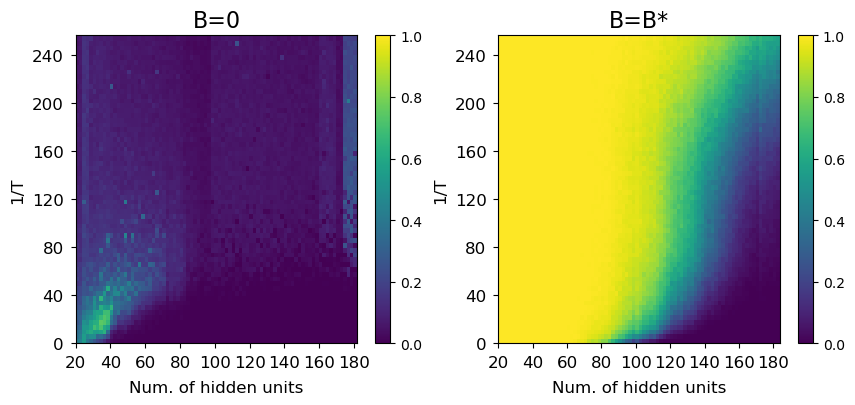}
\caption{Percentage of AIS samples producing a relative error 
lower or equal to $5\%$ with respect to the exact $\log(Z)$ value, as a function
of the number of hidden units and inverse temperature. The left and right panels 
show the results starting from ${\bf B}=0$ and ${\bf B}={\bf B}^*$, respectively.}
\label{fig_B0_Bop_Lokos44}
\end{figure}
\end{center}

In order to test the benefits of using ${\bf B}^*$, we perform several AIS
runs starting from the optimal 
$p^*_0({\bf x}) = 2^{N_h}e^{{\bf B}^*{\bf x}/T}/Z_0$ 
and compare the results to the same calculations starting from the uniform
probability distribution, corresponding to ${\bf B}=0$. As stated above, in
both cases we use $N_\beta=4096$ intermediate chains to obtain $N_s=1024$ AIS
samples.
Figure~\ref{fig_B0_Bop_Lokos44} shows the results obtained in colormap form
for one of the most difficult GWGM cases.
The horizontal axis indicates the number $n_h$ of hidden units
considered, spanning the range from 1 to $N_h=180$, 
obtained by discarding weights
(that is, setting $\omega_{ij}=0$ for $j>n_h$), while 
the vertical axis displays the inverse temperature.
In all cases we use $N_v=20$  visible units, as described in 
Sec.~\ref{sec_Datasets}, thus allowing for the
exact calculation of $\log(Z)$ by {\em brute force}. The maps show the
percentage of the 1024 samples of $\log(Z)$ that differ from the exact value
by less that $5\%$ in each case. 
As it can be readily seen,
the fact that $p_0^*({\bf x})$ is {\em closer} to the RBM probability
distribution 
makes AIS work less and perform better, as expected. Notice, though, that for
some combinations of $T$ and $n_h$ the efficiency of AIS suffers even when
starting from $p_0^*({\bf x})$.
This should not be completely surprising, mostly considering that a mean field
starting probability distribution can still be too far away from that of the
actual RBM, thus indicating that one should look for a different(and unknown)
starting probability distribution.

\begin{center}
\begin{figure}[!t]
\includegraphics[width=0.7\linewidth]{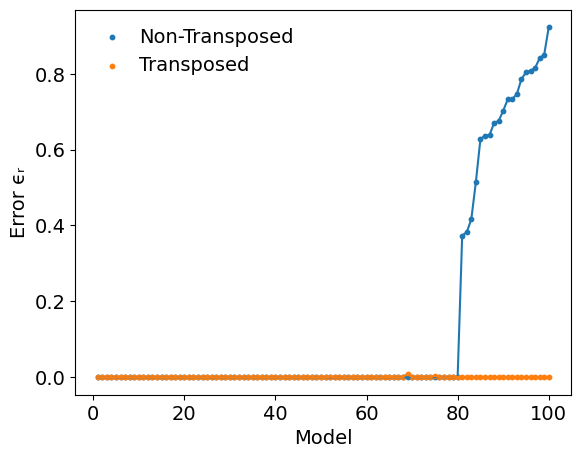}
\caption{Relative error of all models in the transposed and non-transposed
 GWGM datasets, computed as in Eq.~(\ref{relative_error}). For the sake of
 clarity, the models have been sorted according to the relative error of the
 non-transposed results.}
\label{fig_Lokos_Bop_Trans_NonTrans}
\end{figure}
\end{center}

The right panel in Fig.~\ref{fig_B0_Bop_Lokos44} suggests also 
that a mean field starting point can be problematic when the number of hidden
units is much larger than the number of visible ones. This problem is easily
solved noticing that $\log(Z)$ is invariant under the exchange of ${\bf x}$
and ${\bf h}$ in the RBM, associated to replacing the array of weights by its
transpose. 
Based on these results, we have conducted additional tests on the whole GWGM
dataset. In fact, the expectation values $\langle{\bf x}\rangle_n$ can always
be evaluated when the dimension of the hidden space is small, as in the 
present case. It is easy to show that, for binary $[0,1]$ units, one has
\begin{equation}
\langle {\bf x} \rangle_n = 
\sum_{\bf h} p_n({\bf h}) 
\prod_{i=1}^{N_v}
{1 \over 1 + e^{-(b_i + {\bf W}^i {\bf h})/T}} \ ,
\label{xn_optimal_Ex}
\end{equation}
where the sum extends over all hidden states, while $p_n({\bf h})$ and $
{\bf W}^i$ stand for the hidden state probability and the i-th row of the
two-body weights matrix, respectively. Figure~\ref{fig_Lokos_Bop_Trans_NonTrans} 
shows the relative error obtained after averaging ten
repetitions of each AIS run, for 
the 100 GWGM models at $T=1$.
All runs started from ${\bf B}^*$ computed from the exact 
$\langle{\bf x}\rangle_n$, for the transposed and non-transposed configurations.
Results have been sorted in ascending error order of the non-transposed 
configurations in order to get a better view.
As it can be seen, all models are accurately reproduced in the transposed 
case, where the number of hidden units is smaller than the number of visible ones.
On the contrary, about a $20\%$ of the models show large deviations from the 
exact result when the original, non-transposed dataset is evaluated. 
This behavior is similarly observed when performing similar calculations
with the other datasets with large differences in the number of hidden 
and visible units.

\section{Approaching the optimal mean field}
\label{sec_Strategies}

Despite the simplicity of the expressions in Eqs.~(\ref{Bi_xin}) 
and~(\ref {Bi_xin_pm1}), the problem of finding the optimal
${\bf B}^*$
can actually be as hard as finding 
$\log(Z)$ itself, so one has
to devise alternative strategies to approximate it.

Three common strategies are usually employed to face this problem. The
simplest one is to disregard Eqs.~(\ref{Bi_xin}) and~(\ref{Bi_xin_pm1}), 
set ${\bf B}=0$ and sample from the uniform probability distribution,
as discussed above. 
Another common strategy is to set ${\bf B}={\bf b}$ from Eq.~(\ref{energy-RBM})
and to disregard the
contributions of the hidden units,  Despite its simplicity, the resulting
$p_0({\bf x})$ is usually far away from $p_n({\bf x})$. The third approach was
devised in~\cite{Salak_08}
for RBM learning, where $\langle{\bf x}\rangle_n$ is approximated by its
average over the training set. However, this procedure can not be employed
when a training set is lacking, as when dealing with magnetic spin systems for
instance, or when the existing training set does not properly represent the
underlying probability distribution of the system. 

In this work we introduce two alternative strategies to estimate
${\bf B}^*$ that, on the one hand, imply a low computational
cost, and on the other, avoid
some of the drawbacks of the aforementioned choices.
They both rely on finding a suitable approximation to compute 
$\langle {\bf x} \rangle_n$ in Eqs.~(\ref{Bi_xin}) and~(\ref{Bi_xin_pm1}).
At this point many different choices are possible, while keeping in 
mind that none of them will perfectly reproduce the exact 
$\langle {\bf x}\rangle_n$ as we assume the original $p_n({\bf x})$ 
is intractable. 
However, one must keep in mind 
that the resulting probability 
distribution obtained from them is used as the initial point 
for AIS, which will afterwards correct that to produce reliable 
samples of $p_n({\bf x})$.

Among the many possible choices, we introduce the following 
ones:

\begin{itemize}
    
  \item Pseudoinverse (Pinv) approximation: 
    one can look for a state of the complete (visible
    and hidden) space with large probability.
    In this case one works directly with the energy,
    setting to zero the gradients with respect to ${\bf x}$
    of the expression in Eq.~(\ref{energy-RBM}). One then finds
    \begin{equation}
      {\bf x}_p = -({\bf W}^+)^{\rm T} {\bf c} 
      \label{xp_PINV}
    \end{equation}
    where ${\bf W}^+$ is the pseudoinverse of the ${\bf W}$
    matrix. In this work we build ${\bf x}_p$ by rounding 
    the result of Eq.~(\ref{xp_Pinv}) to the $[0,1]$ or the $[-1,1]$
    range, depending on the units used, and approximate
    $\langle {\bf x}\rangle_n$ by ${\bf x}_p$.
    With that we build the corresponding mean-field bias 
    ${\bf B}_{\rm Pinv}$.

  \item Signs from Random Hidden (Signs\_h): 
    The expectation values $\langle {\bf x}\rangle_n$ given in 
    Eq.~(\ref{xn_optimal_Ex}) can only be evaluated when the number
    of hidden units is small, but unfortunately that is 
    not usually the case in real problems. For that reason we
    resort to a heuristic approximation, where a set of 
    hidden states ${\bf h}^{(\alpha)}$ randomly selected from the uniform probability
    distribution is used to get the same number of visible
    states ${\bf x}^{(\alpha)}$ from the conditional probabilities
    $p(x_i^{(\alpha)}=1 | {\bf h}_\alpha) = 
    1 /(1+ e^{-(b_i + {\bf W}^i {\bf h}^{(\alpha)})})$. 
    This expression assigns a probability larger than 0.5 to 
    $x_i^{(\alpha)}=1$ depending on the sign of the argument in the
    exponential.
    Following this, we set the components of ${\bf x}^{(\alpha)}$ 
    to be equal to 1 when $b_i + {\bf W}^i {\bf h}^{(\alpha)} > 0$, and to 0
    in the opposite case. 
    As in most of the calculations performed in this work, we
    build a set of 1024 uniformly sampled ${\bf h}^{(\alpha)}$
    that are used to generate the ${\bf x}^{(\alpha)}$ that are
    finally averaged to get the
    estimation of $\langle{\bf x}\rangle_n$ required to compute the
    approximated bias ${\bf B}_{\rm Signs\_h}$. Notice that this is 
    cost-effective procedure that involves less operations that the 
    pseudoinverse procedure outlined above. This approach is trivially
    extended to $[-1,1]$ units.


\end{itemize}

\begin{center}
\begin{figure}[!t]
\includegraphics[width=0.9\linewidth]{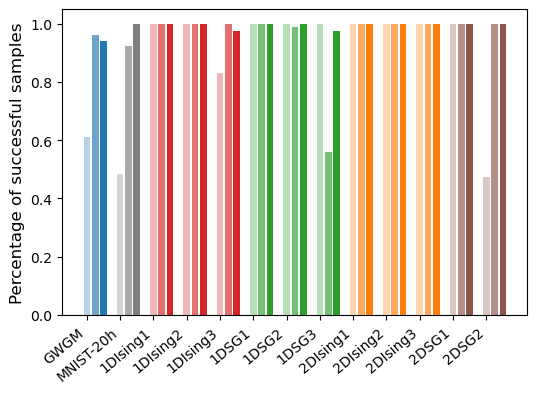}
\caption{Percentage of AIS samples with a relative error lower than $0.05\%$ with respect
to the exact $\log(Z)$ for the different datasets analyzed. 
The left, middle and right bars with different gray levels correspond to the 
predictions starting from ${\bf B}=0$, ${\bf B}={\bf B}_{\rm Pinv}$ and 
${\bf B}={\bf B}_{\rm Signs\_h}$, respectively.}
\label{fig_Percentage_Beq0_PINV_Signsh}
\end{figure}
\end{center}

These two strategies have been used to produce the mean-field probability
distributions of Eq.~(\ref{p0_meanfield}) that are used to start AIS. 
We perform 10 repetitions of each experiment for each model,
producing a total of 1000 final values for the GWGM dataset. 
Figure~\ref{fig_Percentage_Beq0_PINV_Signsh} shows the
statistics obtained for all the datasets, corresponding to the total
amount of AIS predictions producing a relative error of less than $5\%$ 
with respect to the exact value of $\log(Z)$. The lighter, midtone and 
darker bars correspond to ${\bf B}=0$, ${\bf B}={\bf B}_{\rm Pinv}$
and ${\bf B}={\bf B}_{\rm Signs\_h}$, respectively.
As it can be seen, both Pinv and Signs\_h outperforms ${\bf B}=0$ in most
cases, yielding similar results in general. 
It is also worth noticing that for the datasets that do not have 
bias (${\bf b}={\bf c}=0$ in Eq.(\ref{energy-RBM})), ${\bf B}=0$ is
the optimal ${\bf B}^*$ when $[-1,1]$ units are employed.
In this case all three strategies yield very good and similar
results. 

\begin{center}
\begin{figure}[!t]
\includegraphics[width=0.9\linewidth]{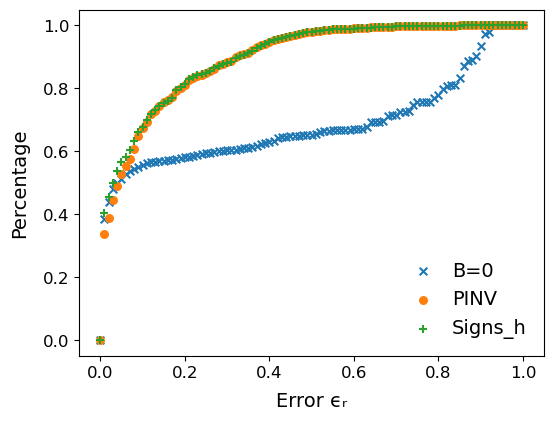}
\caption{Percentage of AIS samples with a relative error lower or equal to $\epsilon_r$
with respect to the exact $\log(Z)$ for the GWGM dataset.}
\label{fig_distrib_samples_Lokos}
\end{figure}
\end{center}

The fact that both ${\bf B}_{\rm Signs\_h}$ and ${\bf B}_{\rm Pinv}$ lead to
overall better AIS predictions than ${\bf B}=0$ is a direct consequence of
the distribution of AIS samples in each case. This is illustrated in
Fig.~\ref{fig_distrib_samples_Lokos} for the GWGM dataset, where all samples
generated from all repetitions of all models have been used to account for a
better statistics. The plot shows the percentage of samples that have a
relative error 
with respect to the exact $\log
(Z)$ equal or lower than $\epsilon_r$, as a function of $\epsilon_r$, for the
${\bf B}=0$, ${\bf B}={\bf B}_{\rm Signs\_h}$ and ${\bf B}={\bf B}_{\rm Pinv}$ 
strategies. As it can be seen, the ${\bf B}=0$ mean field
performs worse than the other two in general, although all three strategies
produce similar results up to $\epsilon \approx 0.05$. For higher values,
though, differences are significant, converging once again towards the end of
the curve where all samples fulfill the condition. In any case, 
we find that ${\bf B}={\bf B}_{\rm Signs\_h}$ and ${\bf B}={\bf B}_{\rm Pinv}$ 
perform very similarly, with minor variations that in the end lead to 
the small prediction differences displayed in 
Fig.~\ref{fig_Percentage_Beq0_PINV_Signsh}. One can thus conclude that, overall, 
the samples generated by ${\bf B}={\bf B}_{\rm Signs\_h}$ 
and ${\bf B}={\bf B}_{\rm Pinv}$ are 
closer to the exact value of $\log(Z)$ than the set produced by ${\bf B}=0$.
Despite that, one could argue that in all cases there is always a large amount 
of samples that fail to predict anything close to the right value.
However, it is worth noticing that this should be the case due to the 
stochastic nature of the AIS algorithm and the exponential way in which 
the generated samples have to be combined, as displayed in 
Eq.~(\ref{log_mean_exp}). Fluctuations above the 
exact value of $\log(Z)$ are exponentially amplified, and have to 
be compensated by a large amount of samples that underestimate its value,
whose contribution is exponentially diminished. We can thus conclude 
that the AIS algorithm has to produce a lot of apparently bad 
samples in order to produce an accurate result. Furthermore, 
this asymmetric generation of samples above and below the exact 
value leads, when not properly balanced, to an underestimation 
of $\log(Z)$, as noticed in~\cite{Burda_2015}.
This picture, though, can be alleviated by increasing the number of
intermediate chains $N_\beta$, at the expense of linearly increasing
the computational cost.

\begin{center}
\begin{figure}[!t]
\includegraphics[width=0.9\linewidth]{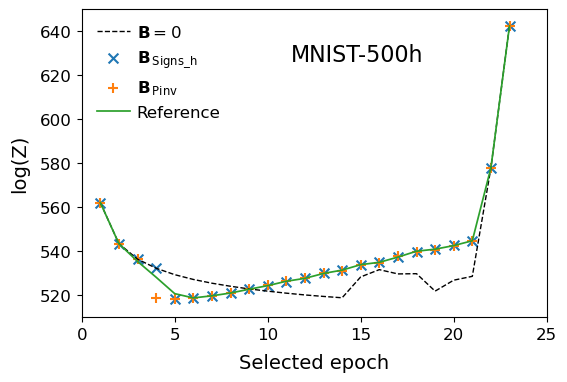}
\caption{Comparison of the AIS estimation of $\log(Z)$ along learning for the
 MNIST dataset with 500 hidden units obtained starting from the different
 mean field probability distributions discussed in this work. The first
 points correspond to the first learning epochs, while the last ones 
 show the predictions obtained at an intermediate learning stage.}
\label{fig_MNIST_500h}
\end{figure}
\end{center}

We finally close the discussion by showing in Fig.~\ref{fig_MNIST_500h} the
value of the partition function estimated with AIS for the  MNIST dataset,
using a RBM model containing $N_h=500$ hidden units. For this system, due to
its large size, there is no exact calculation of $\log(Z)$ and one has to
rely on the predictions obtained employing state-of-the-art techniques found
in the literature. For that matter we take as reference the value obtained
from the procedure outlined in Ref.~\cite{Salak_08}, where the dataset used
to train the RBM is also employed to approximate the mean values required for the
evaluation of ${\bf B}^*$ in Eqs.~(\ref{Bi_xin}) and~(\ref{Bi_xin_pm1}). With
this, we run AIS with $N_s=1024$ and $N_{\beta} = 2^{20}$ to obtain
the reference value (green solid line in the figure). Notice that $N_\beta$ is 
unreasonably large compared to what one would normally use
in order to obtain an as accurate as possible approximation of $\log(Z)$ with
the same number of samples used along this work. The figure shows also the
estimations obtained using ${\bf B}=0$, ${\bf B}={\bf B}_{\rm Signs\_h}$ and
${\bf B}={\bf B}_{\rm Pinv}$ (dotted line, crosses and plus symbols,
respectively). The first 21 points correspond to the first 21 learning epochs
where the RBM weights rapidly evolve, while the last two points correspond to
epochs 40 and 100. As it can be seen, all curves merge at the highest epochs,
while the ${\bf B}=0$ prediction departs from the reference curve at the
early and intermediate epochs. On the contrary, the selected strategies are
hardly distinguishable from the reference line along the whole curve. Despite
the differences between the ${\bf B}=0$ curve and the rest are small, one
should realize that the computational cost involved in using the proposed
strategies is very low, while the prediction obtained are closer to the
reference value. This is something that should be taken into account if the
goal is to get the most accurate but economic prediction of $\log(Z)$.

\section{Summary and Conclusions}
\label{sec_conclusions}

To summarize, 
we have analyzed the performance of the AIS algorithm
in the evaluation of the partition function $Z$ of a Restricted Boltzmann
Machine with a reduced number of samples and intermediate chains. We evaluate
$\log(Z)$ for a number of exactly solvable models which contain a reduced
number of hidden units, as well as in standard physical magnetic spin
problems where the exact value of $\log(Z)$ is also known. In particular we
show that a suitable starting probability distribution $p_0({\bf x})$ of the
mean field form can lead to a big improvement of the AIS estimation of $\log(Z)$ 
for fixed number of samples and intermediate chains. In this scheme, we
build $p_0({\bf x})$ from a RBM with bias terms only corresponding to a local
external magnetic field, and show that the optimal mean field that minimizes
the Kullback-Leibler distance to the RBM probability distribution is directly
related to the averages of the visible states. 
Remarkably, our methodology does not require a training set, and thus it can
be used when none is available. The procedure requires only sampling the RBM.

We also propose two simple strategies to approximate the optimal mean field 
for large systems where the exact averages can not be computed. These
result from a trade-off between simplicity, reduced computational cost, and
accuracy. The first strategy requires the pseudo-inversion of the matrix of
weights, while the second is much cheaper and involves only checking the
signs of a linear transformation of it. Overall, both strategies perform
equal or better than the standard procedure that starts from ${\bf B}=0$ in
the datasets analyzed where $\log(Z)$ is directly accessible. Finally, we
also test them on the MNIST dataset with 500 hidden units, to show that the
estimations obtained are in excellent agreement with the ones from the
procedure outlined in Ref.~\cite{Salak_08}. We expect that the strategies
proposed can be used as the starting point in further studies of $\log(Z)$ in
RBMs with the AIS algorithm, either in isolated form or combined. 


\section*{Acknowledgments}
AP and JM acknowledges financial suppport from 
grant reference PID2021-124297NB-C32 funded by MICIU/
AEI/ 10.13039/501100011033 and by the {\em European Union
NextGenerationEU/PRTR}. FM and JM. thank the Generalitat
de Catalunya for the grant {\em Grup de Recerca SGR-Cat2021
Condensed, Complex and Quantum Matter Group} reference
2021SGR-01411.
FM: This work has been supported by the 
Ministerio de Ciencia e Innovaci\'on
MCIN/AEI/10.13039/501100011033 (Spain) under Grant No. PID2020-113565GB-C21. 
ER: This work was partially supported by MINECO project
PID2022-143299OB-I00 (Spain). 
Part of the hardware used for this research was donated by the
NVIDIA\textsuperscript{\textregistered} Corporation.


\vfill

\pagebreak

\end{document}